\documentclass[aps,pre,twocolumn,superscriptaddress]{revtex4}
\pdfoutput=1
\usepackage{graphicx}
\usepackage{amsmath}
\usepackage{amssymb}
\usepackage{color}
\usepackage{amscd}
\usepackage{bm}  
\usepackage{enumerate}

\DeclareMathOperator{\tanc}{tanc}

\begin{document}
\title{Sheared active fluids: thickening, thinning and vanishing viscosity}

\author{Luca Giomi}
\affiliation{School of Engineering and Applied Sciences, Harvard University, Cambridge, MA 02138,  USA}
\affiliation{Martin A. Fisher School of Physics, Brandeis University, Waltham, MA 02454, USA}

\author{Tanniemola B. Liverpool} 
\affiliation{Department of Mathematics, University of Bristol, Bristol BS8 1TW, U.K.}

\author{M. Cristina Marchetti}
\affiliation{Physics Department \& Syracuse Biomaterials Institute, Syracuse University, Syracuse, NY 13244, USA}

\date{\today}

\begin{abstract}
We analyze the behavior of a suspension of active polar particles under shear. In the absence of external forces, orientationally ordered active particles are known to exhibit a transition to a state of non-uniform polarization and spontaneous flow. Such a transition results from the interplay between elastic stresses, due to the liquid crystallinity of the suspension, and internal active stresses. In the presence of an external shear we find an extremely rich variety of phenomena, including an effective reduction (increase) in the apparent viscosity depending on the nature of the active stresses and the flow-alignment property of the particles, as well as more exotic behaviors such as a non-monotonic stress/strain-rate relation and yield stress for large activities.
\end{abstract}

\maketitle

\section{\label{sec:introduction}Introduction}

\noindent  Colonies of swimming bacteria, \emph{in vitro} mixtures of cytoskeletal filaments and motor proteins, and vibrated granular rods are examples of \emph{active} systems composed of interacting units that consume energy and collectively generate motion and mechanical stresses.  Due to their elongated shape, active particles can exhibit orientational order at high concentration and have been likened to ``living liquid crystals"~\cite{Gruler1999}. Their rich collective behavior  includes nonequilibrium phase transition and pattern formation on mesoscopic scales. It has been modeled by continuum equations built by modifying the hydrodynamics of liquid crystals  to include nonequilibrium terms that account for the activity of the system \cite{TonerRev,SimhaRamaswamySP02,Kruse2004}, or derived from specific microscopic models~\cite{TBLMCM2003,TBLMCMbook}.

A striking property of \emph{confined}  active liquid crystals is the instability of the uniform aligned homogeneous state and the onset of spontaneously flowing states, both stationary and oscillatory~\cite{Voituriez06,GiomiMarchettiLiverpool:2008}. This  occurs  because local orientational order generates active stresses that are in turn balanced by flow, yielding a state that can support local inhomogeneities in the flow velocity and the local alignment, while maintaining a net zero force. Loosely speaking, a confined active liquid crystal ``shears itself'' even in the absence of externally applied forces. It is then not surprising that the rheology of such active liquid crystals in response to an external shear will be very rich. 
\begin{figure}[b]
\centering
\includegraphics[width=.9\columnwidth]{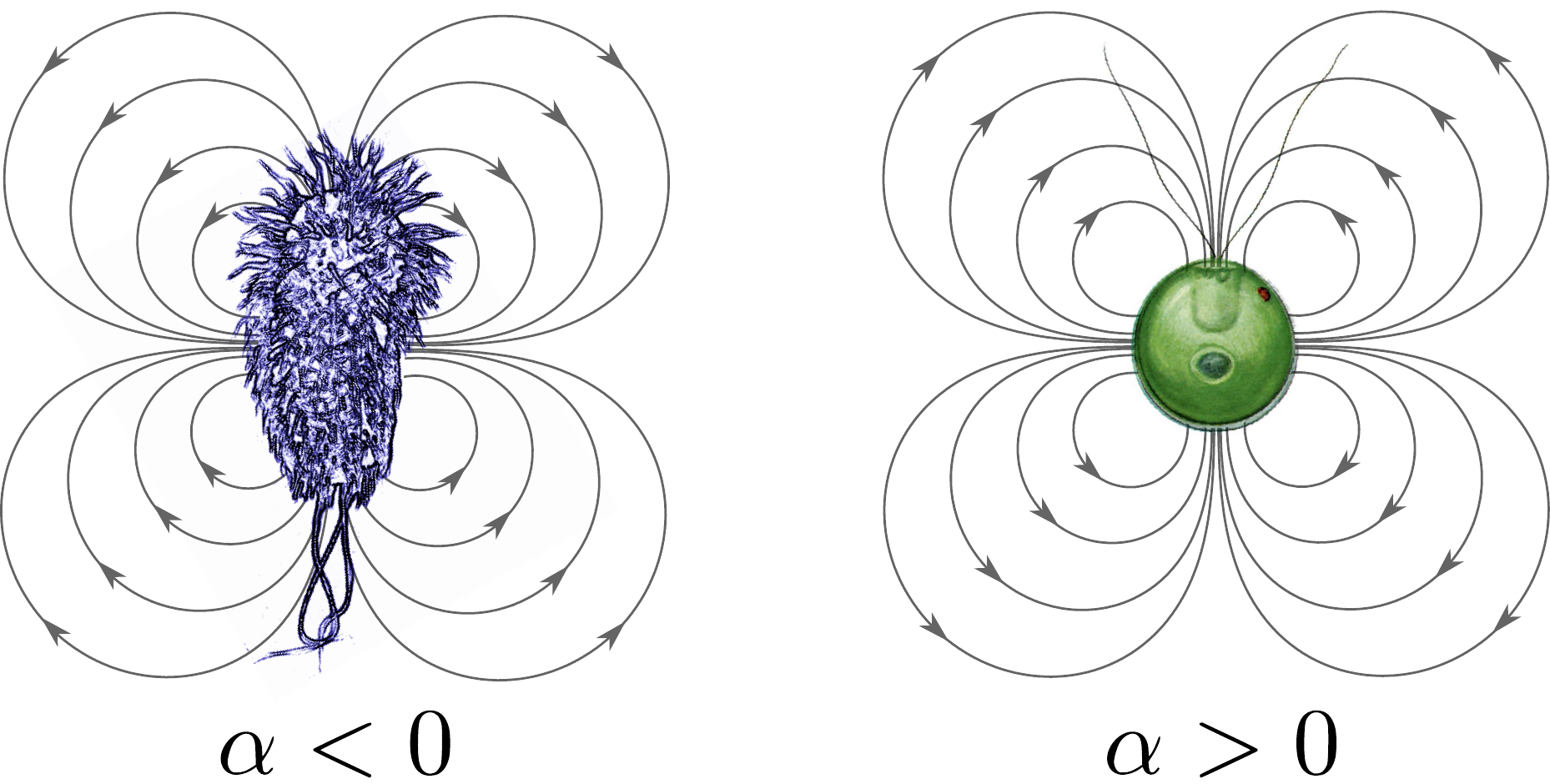}
\caption{\label{fig:bacteria}(color online) Schematic example of the flow field surrounding a tensile (left) and contractile (right) swimming microorganism.}
\end{figure}

Phenomenological work by Hatwalne and collaborators~\cite{Hatwalne04} first pointed out that activity lowers the linear bulk viscosity of tensile suspensions, such as most swimming bacteria, while it enhances the viscosity of contractile systems, and that this enhancement may become very large near the isotropic-nematic transition.  A semi-microscopic model of contractile suspensions of motor-filaments mixtures confirmed these results and predicted an actual divergence of  the viscosity of contractile suspensions  at the  transition~\cite{TBLMCM06}.  Recent numerical studies of active nematic films by Cates \emph{et al}. \cite{CatesEtAl:2008} have confirmed that this result survives when the effect of boundaries is included. In addition, it was found that tensile nematic suspensions can enter a regime of vanishing apparent viscosity in proximity of the isotropic-nematic phase transition. Such a ``superfluid'' window was interpreted by the authors of Ref. \cite{CatesEtAl:2008} as the appearance of bulk shear bands accommodating a range of macroscopic shear-rates at zero stress. Finally, the predicted activity-induced thinning of bacterial suspensions has been demonstrated in recent experiments in \emph{Bacillus subtilis}~\cite{Haines2008,SokolovAranson2009,Haines2009}. 

Active particles exert forces on the surrounding fluid, resulting in local tensile or contractile stresses proportional to the amount of orientational order, $\sigma_{ij}^{\alpha}\sim\alpha n_{i}n_{j}$, where $\alpha$ is proportional to the force exerted by the active particles  on the fluid and ${\bf n}$ a unit vector denoting the direction of broken orientational symmetry. The sign of $\alpha$ determines whether the flow generated by the active particles is tensile ($\alpha<0$) or contractile ($\alpha>0$). In the case of swimming organisms, the former situation describes ``pushers'', i.e., most bacteria (e.g., E. Coli), while the latter corresponds to ``pullers'' (e.g., Chlamydomonas) (see Fig. \ref{fig:bacteria}). 
An important distinction between uniaxial active particles concerns the possibility of forming phases with or without a non-zero macroscopic polarization. Apolar particles are fore-aft symmetric and can form nematic phases in which  macroscopic quantities are invariant for ${\bf n}\rightarrow -{\bf n}$. Polar particles can also form phases characterized by a non-zero macroscopic polarization in the direction of a polar director ${\bf p}$ in which they undergo  collective motion with mean velocity ${\bf v}\sim \beta\,{\bf p}$, with $\beta$ is the typical self-propulsion velocity. This directed motion occurring in polar suspensions contributes to a non-equilibrium local stress  of the form $\sigma_{ij}^{\beta}\sim\beta\,(\partial_{i} p_{j}+\partial_{j} p_{i})$. 

\begin{figure}[t]
\centering
\includegraphics[width=.85\columnwidth]{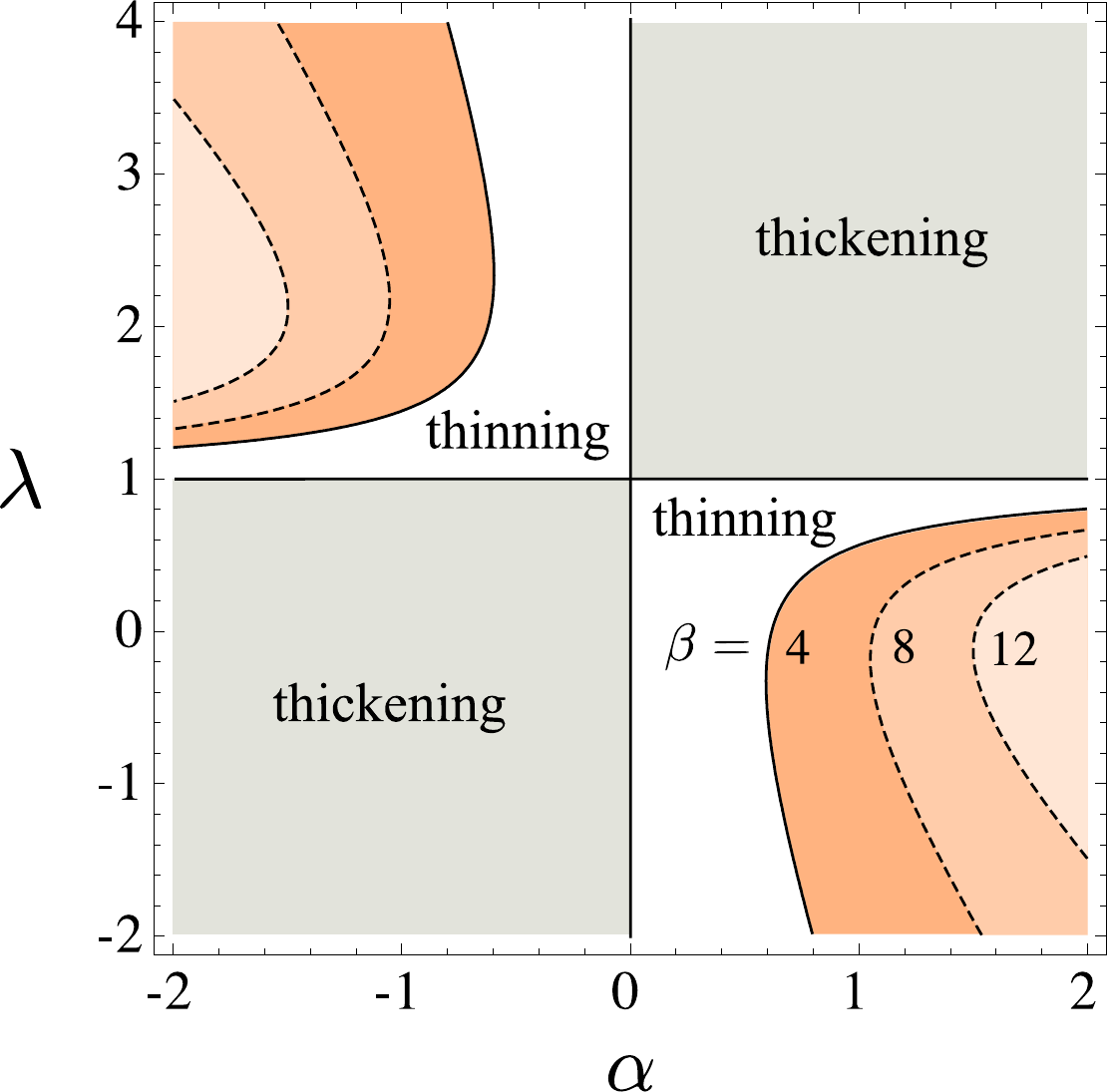}
\caption{\label{fig:alpha-lambda} (color online) The figure displays the regions of parameters where spontaneous flow occurs in an unsheared active film on a substrate. The regions of spontaneous flow are bounded by the critical activity $\alpha_{c1}(\beta)$ given in Eq.~\eqref{alphac1} (solid and dashed lines) and are shaded orange, with lighter shades corresponding to increasing values of $\beta$.  The same critical activity also separates the regions $|\alpha|<\alpha_{c1}$ where the theoretical stress-strain curves are monotonic and the active suspension is either thinned or thickened by activity at small shear rates, as indicated, from the regions $|\alpha|>\alpha_{c1}$ where the theoretical stress-strain curves are nonmonotonic, with possible ``superfluid'' or hysteretic behavior.}
\end{figure}

Most theoretical work has focused on the rheology of active nematic ($\beta=0$), while the shear response of active polar suspensions is far less explored~\cite{Haines2008,Haines2009}.  We find that for a fixed value of $\beta$, the behavior of active suspensions depends on the interplay between the local contractile/tensile stresses, embodied in the parameter $\alpha$, and the flow-aligning behavior of liquid crystalline particles, described by the flow alignment parameter, $\lambda$ \cite{EdwardsYeomans:2009}. Rod-shaped particles typically have $\lambda>0$, spherical particles have $\lambda=0$, while the case $\lambda<0$ describes disk-shaped molecules such as those found in discotic liquid crystals. In passive liquid crystals the magnitude of $\lambda$ controls how the director field responds to a large shear flow away from boundaries. For $|\lambda|>1$ the director tends to align to the flow direction at an angle $\theta_{0}$ such that $\cos 2\theta_{0}=1/\lambda$, while for $|\lambda|<1$ it forms rolls throughout the systems. These regimes are known as ``flow-aligning'' and ``flow-tumbling'' respectively. Understanding of the complex rheology of polar and nematic active suspensions requires exploring the full parameter space, including the important role of boundary conditions. One of the important results of this work is a remarkable exact duality that holds in the regime where the stress-strain relation is linear and shows that tensile ($\alpha<0$) rod-shaped flow-aligning particles ($\lambda>1$) are rheologically equivalent to contractile ($\alpha>0$) discotic flow-tumbling particles ($-1\leq\lambda<0$). Using this result, we present  below a unified description of the linear rheology of  active suspensions of both polar and apolar particles. Some of the results are summarized in the ``phase diagram" of Fig.~\ref{fig:alpha-lambda}. 
This figure shows that the rheological properties of an active film subject to an external shear are closely related to the onset of spontaneous flow in the absence of shear, highlighting the parallel role played in active system by mechanical driving forces, such as a macroscopic strain rate, and internal active driving forces proportional to $\alpha$ and $\beta$. 

An unsheared active film exhibits a transition from the homogeneous aligned state to a ``spontaneously flowing'' state, characterized by spatially  inhomogeneous velocity and director profiles~\cite{Voituriez06}. The transition occurs at a critical activity $\alpha_{c1}$ in a film bounded by one no-slip substrate and a surface that can freely slide, and at a larger value, $\alpha_{c2}>\alpha_{c1}$, in a film bounded by two no-slip planes. The lines separating regions of different shades in Fig.~\ref{fig:alpha-lambda}  are the boundaries $\alpha_{c1}(\beta,\lambda)$ [see Eq.~\eqref{alphac1} below] separating regions of spontaneous flow ($|\alpha|>\alpha_{c1}$) from regions where the homogeneous aligned state is stable ($|\alpha|<\alpha_{c1}$). Interestingly, when the film is subject to an external shear, we find that the flow properties change their {\em qualitative} behaviour at exactly these same critical values of activity. For $\alpha_{c1}<|\alpha|<\alpha_{c2}$,  the theoretical stress-strain rate curves obtained from  our one dimensional model are nonmonotonic (see Fig.~\ref{fig:stress-strain2}) and the active suspension is strongly non-Newtonian. We suggest a number of different interpretations of the nonmonotonic part of the stress-strain rate curve shown in Fig.~\ref{fig:scenarios}. These include macroscopic ``superfluid-like'' behaviour~\cite{CatesEtAl:2008} with zero effective viscosity,  yield-stress behaviour or hysteresis. Finally, for $|\alpha|>\alpha_{c2}$, the theoretical stress-strain curve has a discontinuous jump at zero strain rate, corresponding to a finite ``spontaneous stress''  in the absence of applied shear~\cite{TBLMCM06}. 

\begin{figure}[t]
\centering
\includegraphics[width=0.8\columnwidth]{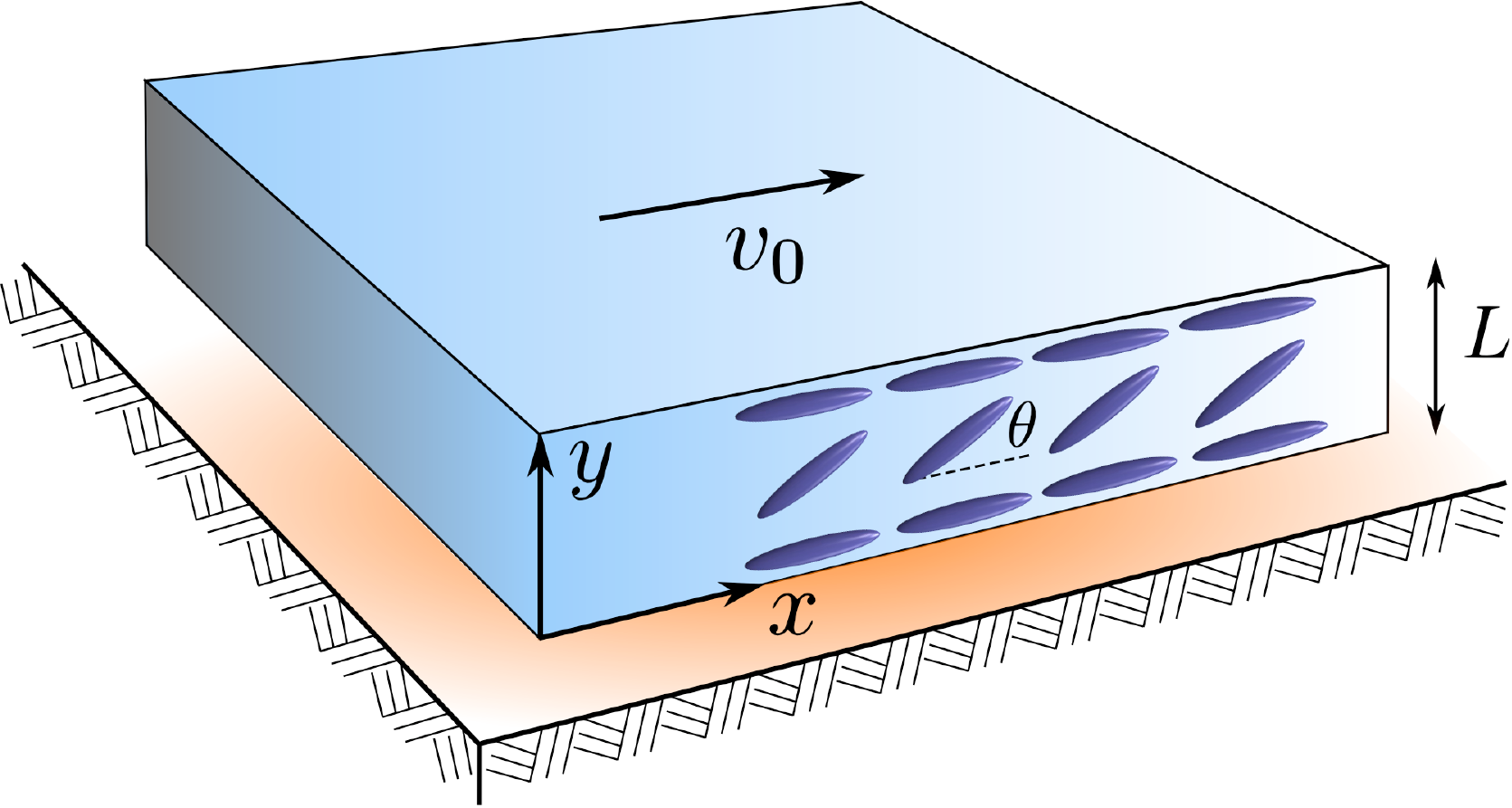}
\caption{\label{fig:sketch}(color online) Schematic representation of a quasi-one-dimensional film of thickness $L$. In our model the film is sitting on a non-slipping surface and is sheared from the top at constant velocity $v_{0}$. The polar rods form an angle $\theta$ with respect to the infinite direction $x$ of the film. Because of the quasi-one-dimensional geometry, the system is invariant for translations along the $x$ axis.}	
\end{figure}

\section{The Model}
Our model of active suspension consists of a two-dimensional film of rod-like particles of length $\ell$ confined to a channel of infinite length along the $x$ axis  and finite thickness $L$ along the $y$ axis (see Fig. \ref{fig:sketch}). Because of the chosen geometry, the system is invariant for translations along the $x$ axis. The total density of the suspension, $\rho=Mc+\rho_{\rm solvent}$, with $c$ the concentration of active particles and $M$ their mass, is assumed to be constant, thus $\nabla\cdot{\bf v}=0$, with ${\bf v}$ the flow velocity.  We assume that the film is sheared at a constant (macroscopic) rate $\dot{\gamma}$ by keeping the lower plate at $y=0$ fixed, while the upper plate at $y=L$ is moved at constant velocity $v_{0}$. The macroscopic shear-rate is defined then as $\dot\gamma=v_{0}/L=\int_{0}^{L}(dy/L)\,u$, where the rate-of-strain tensor $u_{ij}=(\partial_{i}v_{j}+\partial_{j}v_{i})/2$ has only non-zero components $u_{xy}=u_{yx}=\partial_{y}v_{x}/2\equiv u/2$. Theoretical stress-strain curves are obtained by fixing the macroscopic strain rate $\dot\gamma$ and calculating the resulting stress $\sigma$. 

We consider a polarized active suspension and focus only on spatial variations in the direction of the polarization ${\bf P}$. The hydrodynamic equations for an active polar suspension have been formulated by incorporating the active contributions (proportional to the rate of energy  consumed by the active units) into the hydrodynamic equations of a passive polar liquid crystalline film. 
Some of the active contributions, discussed above, are not allowed by the conditions which define liquid crystal systems at equilibrium and hence are {\em intrinsic} to active systems.  Other terms have the same form as those of  passive polar liquid crystals and  can simply be included by modifying the prefactors of the terms obtained from a passive systems.  As such, the {modified} ``passive'' contributions to the equations of motion can be described starting from the non-equilibrium analogue of the Frank free-energy of a suspension of polar particles in a solvent:
\begin{multline*}
F = \int_{\bf r}\,\Big\{
\frac{C}{2}\left(\frac{\delta c}{c_{0}}\right)^{2}+\frac{a_{2}}{2}|{\bf P}|^{2}+\frac{a_{4}}{4}|{\bf P}|^{4}+\frac{K_{1}}{2}(\nabla\cdot{\bf P})^{2}\\[5pt]
+\frac{K_{3}}{2}(\nabla\times{\bf P})^{2}+B_{1}\frac{\delta c}{c_{0}}\,\nabla\cdot{\bf P}+B_{2}|{\bf P}|^{2}\nabla\cdot{\bf P} + \frac{B_{3}}{c_{0}}|{\bf P}|^{2}{\bf P}\cdot\nabla c
\Big\}\,,
\end{multline*}
with $C$ the compressional modulus and $K_{1}$ and $K_{3}$ the splay and bend elastic constant. The parameters $a_i,B_i,K_i,C$ are understood to have both passive and active contributions. In the following we will take $K_{1}=K_{3}=K$. The last three terms in the expression of the free-energy couple concentration and splay and are also present in equilibrium polar suspensions. 

The dynamics of the concentration and the polarization are described by
\begin{subequations}
\begin{equation}
\partial_{t}c = -\nabla\cdot\left[c({\bf v}+c \beta_1 {\bf P})+\Gamma'{\bf h}+\Gamma''{\bf f}\right]\,,\label{eq:c1}\\[-10pt]
\end{equation}
\begin{multline}
[\partial_{t}+({\bf v} +c \beta_2{\bf P})\cdot\nabla]P_{i}+\omega_{ij}P_{j} \\= \lambda u_{ij} P_{j}+\Gamma h_{i}+\Gamma'f_{i}\,,\label{eq:p1}
\end{multline}
\end{subequations}
with $\omega_{ij}=(\partial_{i}v_{j}-\partial_{j}v_{i})/2$ the vorticity tensor, ${\bf h}=-\delta F/\delta {\bf P}$ the molecular field and ${\bf f}=-\nabla(\delta F/\delta c)$. The flow velocity satisfies the Navier-Stokes equation~\footnote{We neglect here convective nonlinearities in the Navier-Stokes equations that are unimportant on the long time scales of interest.}:
\begin{equation}
\rho(\partial_t +{\bf v}\cdot\nabla)v_i=\partial_j\sigma_{ij}\;,
\label{NS}
\end{equation}
with $\nabla\cdot{\bf v}=0$ to guarantee incompressibility, and stress tensor given by dissipative, reversible and active contributions, $\sigma_{ij}=2\eta u_{ij}+\sigma_{ij}^r+\sigma_{ij}^\alpha+\sigma_{ij}^\beta$, with 
\begin{subequations}
\begin{gather}
\sigma_{ij}^\alpha=\frac{\alpha c^{2}}{\Gamma} \big(P_iP_j+\delta_{ij}\big)\;,\label{sigma-a}\\[5pt]
\sigma_{ij}^\beta=\frac{\beta_3 c^{2}}{\Gamma}\big[\partial_iP_j+\partial_jP_i+\delta_{ij}\bm\nabla\cdot{\bf P}\big]\;,\label{sigma-b}\\[5pt]
\sigma_{ij}^r=-\Pi\delta_{ij}-\frac{\lambda}{2}(P_ih_j+P_jh_i)+\frac{1}{2}(P_ih_j-P_jh_i)\;,\label{sigma-r}
\end{gather}
\end{subequations}
%
%TBL
where $\Pi$ is the pressure,  $\eta$ the shear viscosity, and
we have assumed an isotropic viscosity tensor. 
We now consider a solution deep in the polarized state and neglect fluctuations in the magnitude of the polarization, i.e., assume $|{\bf P}|=\sqrt{-a_{2}/a_{4}}$. For simplicity we also redefine units so that $|{\bf P}| = 1$. The condition ${\bf P} = {\rm constant}$ determines the longitudinal part $h_{\parallel} = {\bf p}\cdot{\bf h}$ of the molecular field that can then be eliminated from the hydrodynamic equations. The details associated with imposing the constancy of the magnitude of the polarization and deriving the hydrodynamic equations solely in terms of the polar director ${\bf p} ={\bf P}/|{\bf P}|$  are given in Appendix \ref{sec:appendix}. With this choice, the hydrodynamic equations for ${\bf p}$ and $c$ can be written in the form
%TBL
%
\begin{subequations}
\begin{equation}
\partial_tc+\bm\nabla\cdot c({\bf v}+\beta_1 c {\bf p})=\partial_i\left[D_{ij}\partial_j c+\lambda \gamma' u_{kl}p_{k}p_{l}p_{i}\right]\;,\label{c}\\[-7pt]
\end{equation}
\begin{multline}
[\partial_t+({\bf v}+\beta_2 c {\bf p})\cdot\bm\nabla]p_i+\omega_{ij}p_j\\[5pt]
=\delta_{ij}^T\left[\lambda u_{jk}p_k+\frac{w}{c_{0}}\partial_ic-\frac{\gamma'w}{c_{0}}\partial_{j}\nabla\cdot{\bf p}+\kappa\nabla^{2} p_{j}\right]\;,\label{p}
\end{multline}
\end{subequations}
with $\gamma'=\Gamma'/\Gamma$, $\kappa=\Gamma K$, $w=\Gamma(B_{1}-B_{3})$ and $\delta_{ij}^{T}=\delta_{ij}-p_{i}p_{j}$ the transverse projection operator. $D_{ij}$ is an effective diffusion tensor given by
\begin{equation}\label{eq:diffusion-tensor}
D_{ij} = D_1\delta_{ij}+D_2p_{i}p_{j}\;,
\end{equation}
where $D_1=D-\gamma'w/c_0$ and $D_2=\gamma'w/c_0-D\xi$. 
Finally, the reversible part of the stress tensor $\sigma_{ij}^{r}$ becomes:
\begin{align*}
\sigma_{ij}^{r} 
&= -\delta_{ij}\Pi + \lambda p_{i}p_{j}p_{k}\left[\frac{w}{c_{0}\Gamma}\,\partial_{k}c+K\nabla^{2}p_{k}\right]\\
&-\frac{\lambda}{2}\left[\frac{w}{c_{0}\Gamma}(p_{i}\partial_{j}c+p_{j}\partial_{i}c)+K(p_{i}\nabla^{2}p_{j}+p_{j}\nabla^{2}p_{i})\right]\\[2pt]
&+\frac{1}{2}\left[\frac{w}{c_{0}\Gamma}(p_{i}\partial_{j}c-p_{j}\partial_{i}c)+K(p_{i}\nabla^{2}p_{j}-p_{j}\nabla^{2}p_{i})\right]\\
&-\lambda\Gamma'\xi\,p_{i}p_{j}(Dp_{k}\partial_{k}c+wp_{k}\partial_{k}\partial_{l}p_{l})
+\frac{\lambda^{2}}{\Gamma} p_{i}p_{j} u_{kl} p_{k}p_{l}\,.
\end{align*}
The equations for an active suspension have been written down phenomenologically and also derived from various semi-microscopic models. The structure of of the equations is generic and applies to a broad class of ``living liquid crystals''. The parameters in the equations are of course system and model specific.   In motor/filament mixtures activity arises from clusters of motor proteins crosslinking pairs of filaments. The active couplings are therefore of order $c^2$ in this case~\cite{TBLMCM2003,TBLMCMbook}. In suspensions of swimming microorganisms, activity can be described in terms of the active force $f$ that each swimmer exerts on the surrounding fluid. In this case the active couplings arise even at the single-swimmer level and are of order $c$~\cite{BaskaranMarchetti:2009}. Estimates for the active parameters obtained from semimicroscopic models are summarized in Table~\ref{table1}.
\begin{table}
\centering
\begin{ruledtabular}
\begin{tabular}{crr}
$\quad$ & filaments/motors ($\sim$) & swimmers ($\sim$) \\
\hline
$\beta_1$ & $\tilde{m} u_0\ell^2 \qquad\qquad$ & $ v_{\rm sp}/c       \qquad$\\
$\beta_2$ & $-\tilde{m} u_0\ell^2\qquad\qquad$ & $ v_{\rm sp}/c       \qquad$\\
$w$       & $\tilde{m} u_0\ell^2 \qquad\qquad$ & $-v_{\rm sp}/c       \qquad$\\
$\alpha$  & $\tilde mu_{1} \ell^2\qquad\qquad$ & $ f\ell^3/(\zeta c)  \qquad$\\
$\beta_3$ & $\tilde mu_{0} \ell^2\qquad\qquad$ & $ v_{\rm sp}/c       \qquad$\\
\end{tabular}
\end{ruledtabular}
\caption{Estimates of active parameters for two types of active suspensions: (i) mixtures of cytoskeletal filaments and cross-linking motor proteins~\cite{TBLMCM2003,TBLMCMbook,defnote}, with $\tilde{m}$ a dimensionless density of crosslinking motor clusters , $u_0$ the speed at which motor proteins walk on filaments, in turn proportional to the rate of ATP consumption, and $|u_1|\sim u_0\ell_m$, with $\ell_m$ the size of a motor cluster; and (ii) swimming microorganisms~\cite{BaskaranMarchetti:2009}, where $v_{\rm sp}\sim (f/\zeta)\,\epsilon$ is the self-propulsion speed of an individual organisms, with $f$ the force that swimmers exert on the fluid, $\epsilon<1$ a dimensionless number determined by the shape of the swimmer and $\zeta\sim 1/\Gamma$ is the longitudinal friction coefficient of a rod-like swimmer of length $\ell$. For both systems the precise values of parameters obtained from each microscopic model differ from the above by numericsl constants of order unity.}
\label{table1}
\end{table}
The equations for an active nematic can be obtained from those of a polar systems by setting $\beta_i=w=0$. In the following we assume $\beta_1=\beta_3=-\beta_2=\beta$, as appropriate for motor filament-systems.

It is convenient to work with dimensionless quantities. Spatial variables are normalized with the length $\ell$ of the rods. Thus $y\rightarrow y/\ell$. Temporal variables are normalized with the time scale of splay and bending fluctuations, thus $t\rightarrow t/\tau$ where $\tau=\ell^{2}/\kappa$.  A mass scale is set by $\tau/\Gamma$. All the other quantities are normalized accordingly. In these units the hydrodynamic equations for the  rods concentration $\phi=c/c_0$, with $c_0$ the mean density, and the director/polarization angle $\theta$, with ${\bf p}=(\cos\theta,\sin\theta)$, for the geometry of interest are 
\begin{subequations}\label{eq:hydrodynamic-equations}
\begin{gather}
\rho(\partial_{t}+v_{y}\partial_{y})v_{x} = \partial_{y}\sigma_{xy}\label{eq:v}\\[7pt]
\partial_t \phi 
= \partial_{y}\big\{\beta \phi^{2}\sin\theta+{\cal D}(\theta)\partial_y \phi
+\lambda u\sin\theta\sin2\theta\big\}\,,\label{eq:phi}\\[7pt]
\partial_t{\theta} 
= -\beta \phi\sin\theta\partial_y\theta+w\cos\theta\partial_y\phi+{\cal K}(\theta)\partial_y^2\theta \notag\\
+ w\cos\theta\sin\theta (\partial_y\theta)^2 -u(1-\lambda\cos 2\theta)\,, \label{eq:theta}
\end{gather}
\end{subequations}
where ${\cal D}(\theta)=D(1-\xi\sin^2\theta)-w\cos^2\theta$ is a  diffusion coefficient, ${\cal K}(\theta)=1-w\cos^2\theta$ describes the energy cost of bend and splay deformations, and $\lambda$ is the flow-alignment parameter. In a steady state the  stress tensor  $\sigma_{xy} \equiv\sigma$ is constant across the film and it is given by
\begin{align}\label{eq:u} 
\sigma &= u\Big[\eta+\lambda^{2}\sin^{2}2\theta\Big] 
+\lambda w\sin^2\theta\sin 2\theta(\partial_y\theta)^2\notag\\[3pt]
&\quad+[w-\lambda w_{0}-\lambda(w-w_{0})\cos 2\theta]\cos\theta\partial_y\phi\notag\\[5pt]
&\quad+\alpha\phi^{2}\sin2\theta-2\beta\phi^{2}\sin\theta \partial_y\theta
\,, 
\end{align}
with $\eta$ the bare viscosity and $w_{0}$ a constant proportional to the ratio between the translational and orientational diffusion coefficients (i.e. $w_{0}\sim D/K$). 
Our goal is to study the relation between the induced shear stress $\sigma$ and the applied shear rate $\dot{\gamma}$ as a function of the two fundamental active parameters $\alpha$ and $\beta$ representing the magnitude of the internal contractile/tensile stress and the velocity scale of directed motion. In order to construct a $\sigma$ vs $\dot{\gamma}$ map, we integrate Eqs. \eqref{eq:hydrodynamic-equations} numerically with boundary conditions $v_{x}(0)=0$ and $v_{x}(L)=v_{0}$, $\theta(0)=\theta(L)=0$ and $j_{y}(0)=j_{y}(L)=0$ which implies $\phi'(0)=\phi'(L)=0$. As initial conditions we choose $\theta(y,0)=0$ and $\phi(y,0)=1$. 

In the absence of applied shear, active polar and nematic  films exhibit a transition from a quiescent ($v_x=0$) aligned ($\theta=0$) state to a state of spontaneous flow, with both inhomogeneous alignment and velocity profiles. The critical value of activity where the instability occurs depends on boundary conditions. For a film bounded by a no-slip substrate and a surface that can freely slide it is given by \cite{GiomiMarchettiLiverpool:2008}:
\begin{equation}
\label{alphac1}
\alpha_{c1}(\beta,\lambda)
=\left(\frac{\pi}{L}\right)^{2}\frac{\eta(1-w)}{2\phi_{0}^{2}(1-\lambda)}
+\frac{\beta w[\eta+(1-\lambda)^{2}]}{2(1-\lambda)(D-w)}\,,
\end{equation}
and the spontaneously flowing state has $\sigma=0$. For a film bounded by two no-slip surfaces the critical value is $\alpha_{c2}=4\alpha_{c1}$ and the spontaneously flowing state is characterized by a finite value of $\sigma$. The regions of spontaneous flow  in the $(\lambda,\alpha)$ plane are displayed in shades of orange in Fig.~\ref{fig:alpha-lambda}. In these regions the film exhibits strongly nonlinear rheology, with nonmonotonic stress-strain curves, as described below. 

\begin{figure}[t]
\centering
\includegraphics[width=1.\columnwidth]{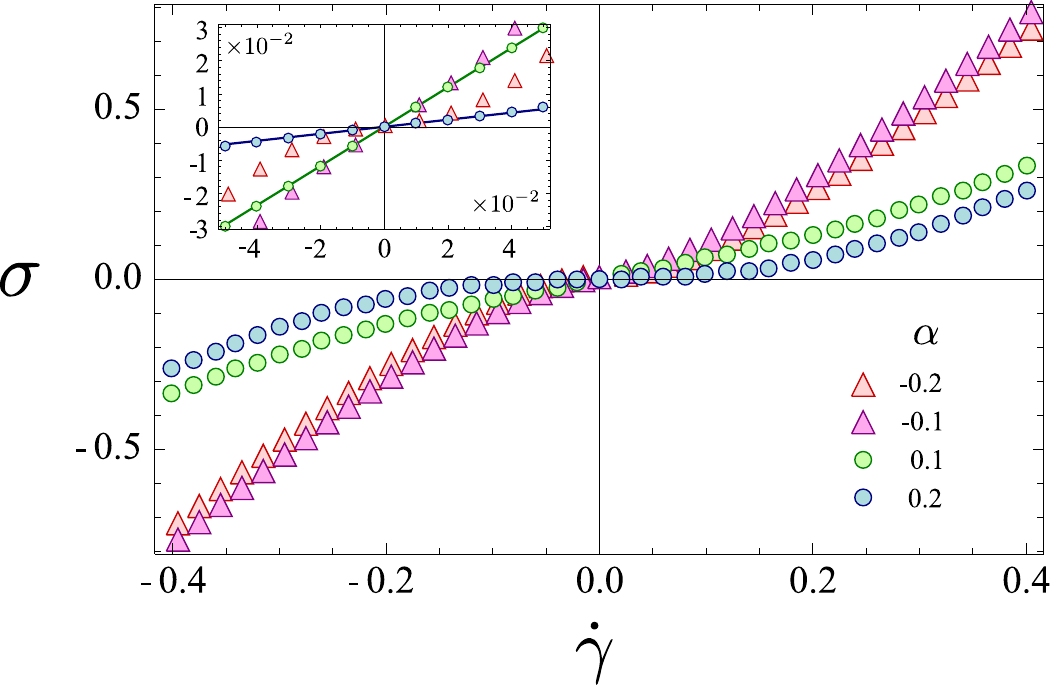}
\caption{\label{fig:stress-strain1} Stress ($\sigma$) vs strain ($\dot{\gamma}$) for an active nematic ($\beta=w=0$) suspension for various $\alpha$. Flow-tumbling system with $\lambda=0.1$ are marked by circles and flow-aligning systems with $\lambda=1.9$ by triangles. Other parameters are set $L/\ell=5$, $\eta=1$, $\phi_0=1$, $D=1$ and $\xi=0.3$. The inset shows the comparison with the analytical result given in Eq.~\eqref{eq:eta-eff}.}
\end{figure}

\section{Linear rheology of weakly active systems}

\begin{figure}[t]
\centering
\includegraphics[width=.9\columnwidth]{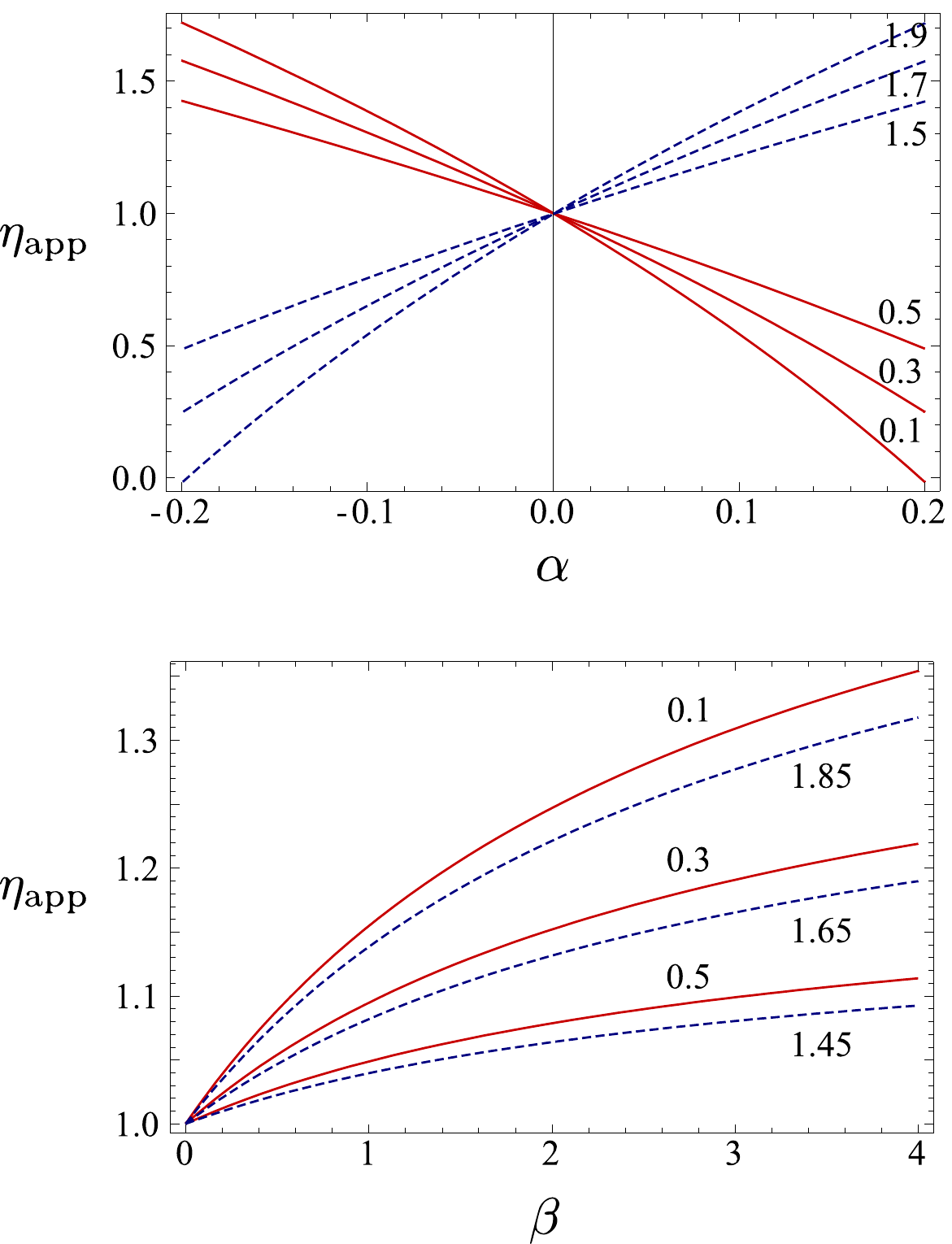}
\caption{\label{fig:eta-apparent} (color online) Apparent viscosity $\eta_{\rm app}$ for active nematic (top) and polar (bottom) suspensions. Solid/red lines represent flow-tumbling systems ($\lambda<1$) while dashed/blue lines flow-aligning systems ($\lambda>1$). The corresponding values of $\lambda$ are indicated next to the lines.  In the bottom plot $\alpha$ was set to zero. The top frame emphasizes the duality discussed in the text.}
\end{figure}

\noindent For $|\alpha|<\alpha_{c1}$, corresponding to the gray regions of Fig.~\ref{fig:alpha-lambda}, the stress strain curves are monotonic and remain linear over a broad range of $\dot\gamma$, as shown in Fig.~\ref{fig:stress-strain1}. Non-Newtonian behavior sets in at smaller values of $\dot\gamma$ with increasing $\alpha$.
As the value of $\alpha$ is increased the slope of the linear portion of the stress-strain curves for $\alpha<\alpha_{c1}$  decreases with increasing $\alpha$, indicating that contractile active stresses  lower the effective viscosity of the system.  The effective linear  viscosity can be calculated analytically by solving  Eqs. \eqref{eq:theta} and \eqref{eq:phi} perturbatively in $\sigma$ by expanding the fields $\theta$ and $\phi$ as $\theta = \theta_{0}+\sigma\theta_{1}+\sigma^{2}\theta_{2}\ldots$ and $\phi = \phi_{0}+\sigma\phi_{1}+\sigma^{2}\theta_{2}\ldots$ The quantities $\theta_{0}$ and $\phi_{0}$ represents here the stationary solution of the hydrodynamic equations in absence of shear flow. If the suspension is in an aligned state at $t=0$, when the shear is switched on, then $\theta_{0}=0$ and $\phi_{0} = \text{const}$. We note, however,  that  this perturbation analysis breaks down in the region $\alpha>\alpha_{c1}$ of spontaneous flow, as in that case both $\theta,\phi$ are spatially varying even at $\sigma=0$. It is straightforward to solve Eqs. \eqref{eq:theta} and \eqref{eq:phi} to first order in $\sigma$. We then obtain the linear apparent viscosity defined as  $\eta_{\rm app}=\lim_{\dot\gamma\rightarrow 0}\sigma /\dot{\gamma}$ and given by
\begin{equation}\label{eq:eta-eff}
\eta_{\rm app}=\frac{\eta(1+\zeta)}{\zeta+\tanc\left(\frac{kL}{2}\right)}\,,
\end{equation}
where $\tanc(x) = \tan(x)/x$ and 
\begin{subequations}\label{eq:k}
\begin{gather}
\zeta = \frac{\eta w \beta}{(1-\lambda)[\beta w(1-\lambda)-2\alpha(D-w)]}\,,\\[10pt]
k^{2}=\frac{2\alpha\phi_{0}^{2}(1-\lambda)}{\eta(1-w)}-\frac{\beta w\phi_{0}^{2}}{(1-w)(D-w)}\left[1+\frac{(1-\lambda)^2}{\eta}\right]\,,
\end{gather}
\end{subequations}
For passive system $\alpha=\beta=w=0$, and $\eta_{\rm app}=\eta$, as expected. For active nematic, $\beta=w=0$ and  the apparent viscosity is  simply
\begin{equation}\label{eq:eta-nematic}
\eta_{\rm app} = \frac{\eta}{\tanc\left(\frac{k}{2}\frac{L}{\ell}\right)}\,,
\end{equation}
with  $k=\sqrt{2\alpha\phi_{0}^{2}(1-\lambda)/\eta}$.   If $\alpha(1-\lambda)<0$, $k$ is imaginary and the $\tan$ function at the denominator of $\eta_{\rm app}$ is replaced by its hyperbolic counterpart. Since $\tanh(x)$ increases more slowly than $x$, the resulting apparent viscosity will increase. If $\alpha(1-\lambda)>0$, $k$ is real and since the $\tan(x)$ function grows more rapidly than $x$ we expect then a rapid decrease in the apparent viscosity as $|\alpha|$ is increased. 
This shows that the linear rheology of pullers/contractile systems with $\lambda<1$ are the same as those of pushers/tensile systems with $\lambda>1$. From Eq. \eqref{eq:eta-eff} it is indeed simple to prove that the apparent viscosity $\eta_{\rm app}$ is invariant under the transformation 
\begin{equation}\label{eq:eta-invariance}
\eta_{\rm app}(\alpha,\beta,\lambda) = \eta_{\rm app}(-\alpha,\beta,2-\lambda)\,.
\end{equation}
\begin{figure}[b]
\centering
\includegraphics[width=.9\columnwidth]{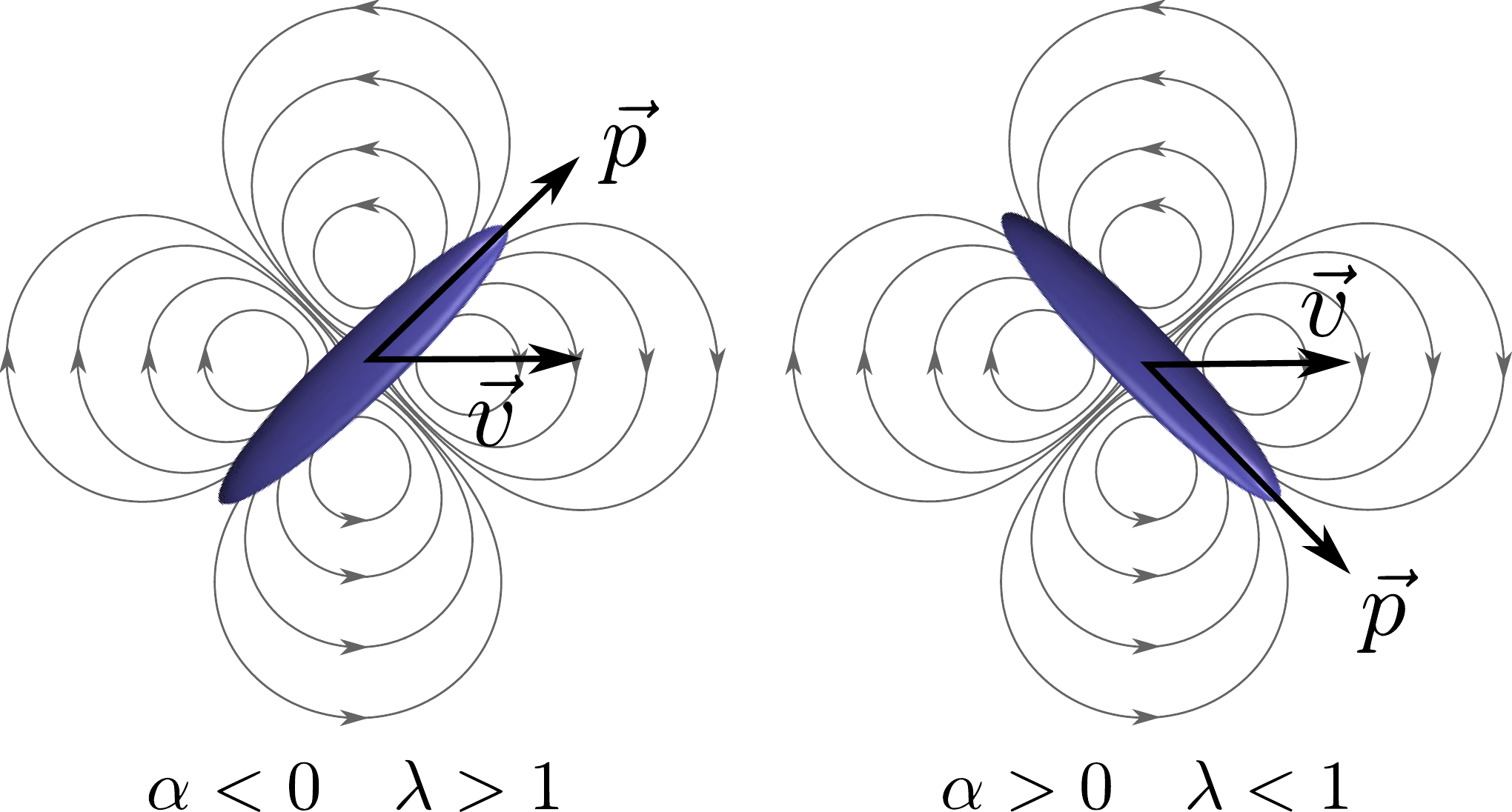}
\caption{\label{fig:active-flow}Schematic example of the flow field surrounding a tensile/flow-aligning (right) and contractile/flow-tumbling active particle. For the choice of the parameters $\alpha$ and $\lambda$ given in Eq. \eqref{eq:eta-invariance} the two flows are identical, leading to an equal apparent viscosity.}
\end{figure}%
Thus  flow-aligning pullers  with $\lambda=1+\epsilon$ (for $0\le\epsilon<1$) will exhibit the same apparent viscosity of ow-tumbling pushers with $\lambda=1-\epsilon$: $\eta_{\rm app}(-|\alpha|,\beta,1+\epsilon)=\eta_{\rm app}(|\alpha|,\beta,1-\epsilon)$. This duality is displayed in the top frame of Fig.~\ref{fig:eta-apparent} that shows the linear apparent viscosity of active nematic suspensions  as a function of $|\alpha|$  for several values of $\lambda$.  The solid curves (red online) show that both contractile/flow tumbling suspensions and tensile/flow aligning ones are thinned by activity. The dashed curves (blue online) refer to either contractile/flow aligning suspensions or tensile/flow tumbling ones and show that these systems are thickened by activity.  
%%TBL
Bacteria such as E-Coli are pushers ($\alpha<0$) and generally elongated in shape, corresponding to $\lambda>1$. Our results therefore confirm the activity-induced thinning of bacterial suspensions first predicted by Hatwalne et al~\cite{Hatwalne04} and recently observed in ~\cite{SokolovAranson2009}. In contrast,  algae like Chamydomonas that propel themselves from the front (and are therefore pullers, with $\alpha>0$). Whether they are thickened or thinned by activity  depends intimately on their shape, i.e. on whether they can be described as objects with  $\lambda>1$ or $\lambda < 1$.  Similarly, motor/filament mixtures are generally contractile ($\alpha>0$) are are expected to be thickened or thinned by activity depending on the effective value of $\lambda$.
%%TBL

This duality has a simple interpretation. Active contractile (tensile) particles produce an ingoing (outgoing) flow in the surrounding fluid, but while flow-aligning particles orient at a positive angle with respect to the flow direction, flow-tumbling particles orient at a negative angle under a small applied shear (see Fig. \ref{fig:active-flow}). As a result, the average flow fields produced in the surrounding fluid are identical in the two cases and produce the same resistance to the imposed shear flow. This equivalence holds only for small applied shear stresses. For large shear-rates the configuration of the director field of a flow-tumbling suspension is dramatically different from that of flow-aligning one and the similarity between the two flow-fields no longer holds.

\section{Nonlinear rheology of strongly active systems}

\begin{figure}[t!]
\centering
\includegraphics[width=.9\columnwidth]{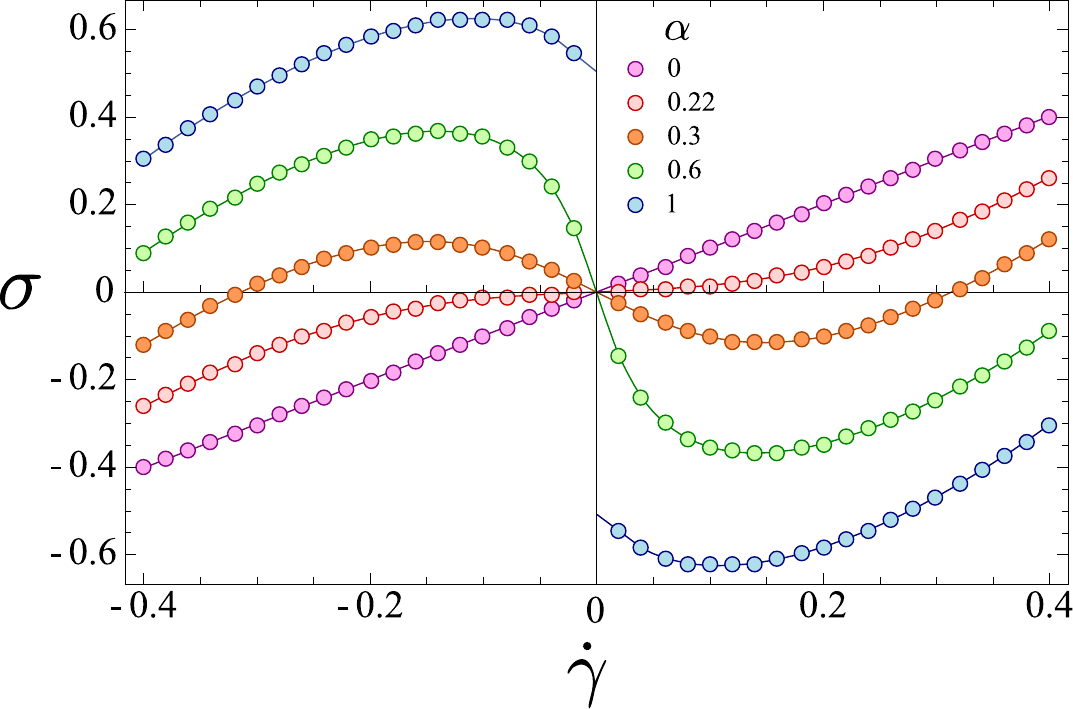}
\caption{\label{fig:stress-strain2} Stress-strain curves of a nematic suspension ($\beta=w=0$) obtained by numerical solution of the active hydrodynamic equations for several values of $\alpha$. $\alpha_{c1}=0.219$ and $\alpha_{c2}=0.877$ for the parameters chosen in the numerical solution.}
\end{figure}

\noindent The linear apparent viscosity given by Eq. \eqref{eq:eta-nematic} vanishes at $\alpha=\alpha_{c1}$, suggesting the onset  of  a superfluid-like behaviour above this critical value of activity~\cite{CatesEtAl:2008}. For $\alpha>\alpha_{c1}$, { the linearized approximation breaks down} and  the stress versus (average) strain rate curve obtained by numerical solution of the equations is  nonlinear and nonmonotonic, as shown in Fig.~\ref{fig:stress-strain2}. We emphasize that the flow profiles are always 
inhomogeneous with varying velocity gradients and director orientation.
For $\alpha_{c1}<\alpha<\alpha_{c2}$ the theoretical stress versus macroscopic (average) strain rate curve goes through the origin and exhibits a region of negative $d\sigma/d\dot\gamma$, that would in principle be mechanically unstable. 
What would be measured in an experiment would, however,  depend critically on details of the experimental procedure and the particular apparatus. 
To study the steady state rheology there are in general  two natural classes of experiments: either (i) one tunes the stress $\sigma$ and measures the resulting strain rate $\dot\gamma$ or (ii) one does a sweep through the values of strain rate $\dot\gamma$ and measures the stress $\sigma$.   If the stress-strain rate curve is monotonic, the two procedures are expected to yield the same result. However, this is no longer the case as soon as the response exhibits nonmonotonicity.

\begin{figure}[b]
\centering
\includegraphics[width=1.\columnwidth]{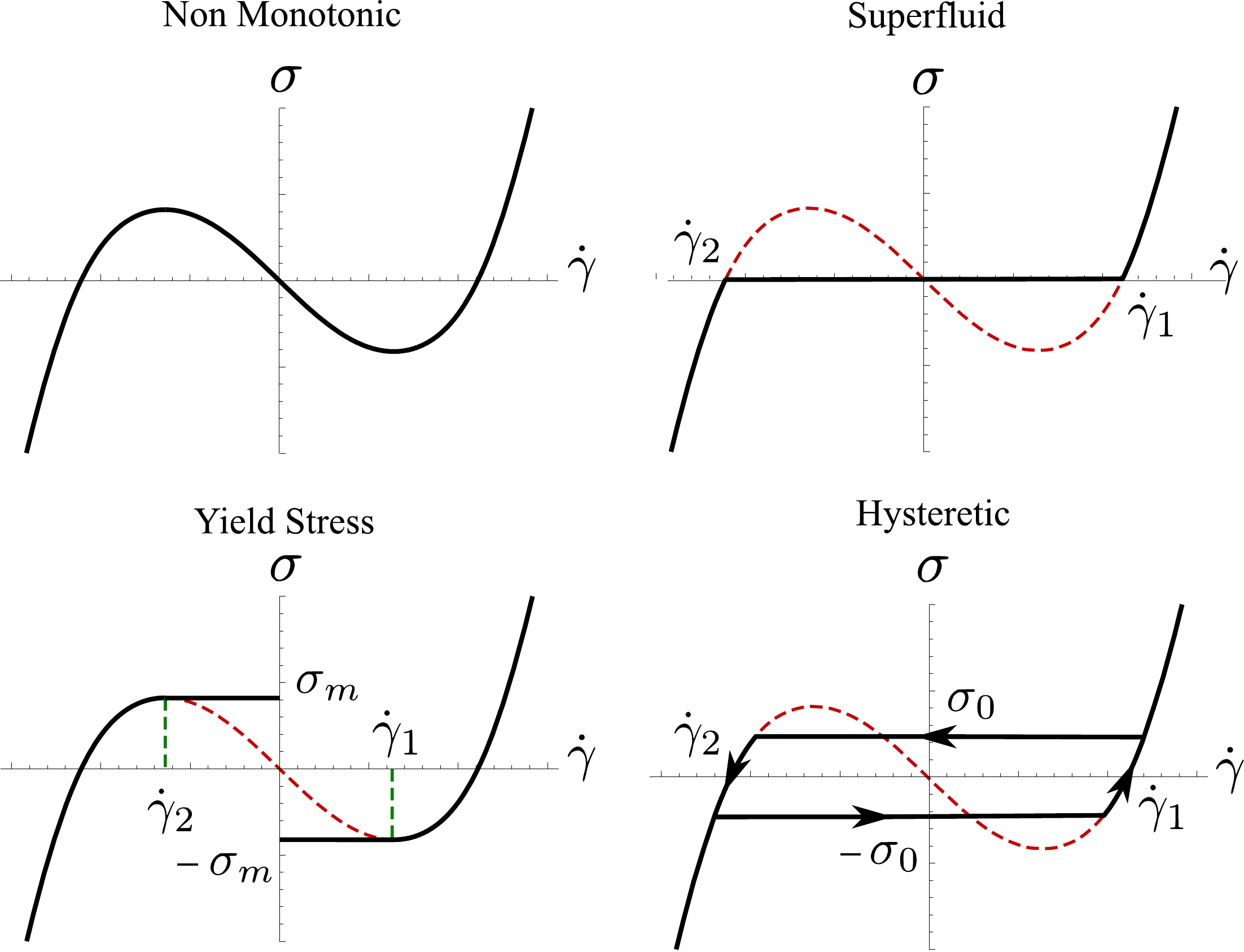}
\caption{\label{fig:scenarios}The top left frame display a typical theoretical stress-strain curve of a nematic active suspension in the region $\alpha_{c1}<|\alpha|<\alpha_{c2}$. The theoretical curve is obtained by tuning $\dot\gamma$ and calculating the resulting $\sigma$  and exhibits a region of $d\sigma/d\dot\gamma<0$. The other three frames show three possible  experimental stress-strain curves obtained by tuning $\sigma$ and measuring $\dot\gamma$ that could be consistent with the theoretical curve. The top right frame displays the ``superfluid'' scenario suggested in~\cite{CatesEtAl:2008}, with bulk shear bands accommodating different macroscopic shear rates and zero net stress, so that the apparent viscosity of the system is simply zero. The bottom left frame  shows a yield-stress like behaviour with a yield stress $\sigma_y=\sigma_m$. The last scenario is described in the bottom right frame and corresponds to a hysteretic stress-strain curve where the suspension can  accommodate a range of macroscopic strain rates maintaining  a constant total stress $\pm\sigma_0$.}
\end{figure}

%%TBL
An important question, then, is what is the shape of the stress-strain rate curve that would be obtained experimentally for $\alpha>\alpha_{c1}$ in an experiment where one tunes the {\em macroscopic} strain rate $\dot\gamma$. Several scenarios are possible, as shown in Fig.~\ref{fig:scenarios} for a non-monotonic curve with maximum/minimum at $\pm \sigma_{m}$.
 
(i) One scenario, suggested  recently~\cite{CatesEtAl:2008} based on numerical studies in the proximity of the isotropic-nematic phase transition and for small value of the active stress $\alpha$ is the appearance of bulk shear bands accommodating a range of macroscopic shear-rates at zero stress. This would correspond to the bulk stress-strain curve displayed in the top right frame of Fig.~\ref{fig:scenarios} and characterized as ``superfluid'' behavior. In the simplest picture the sheared suspension would separate in bands of constant and opposite strain rates, each with zero stress. 
For the systems studied here (deep in the ordered phase, either nematic or polar), we find that the equations of motion provide no mechanism for selecting a particular value of the stress plateau and are unable to find a stable stress-plateau at any value of $|\sigma|< \sigma_m$ (including $\sigma=0$, see Fig. \ref{fig:scenarios}). Furthermore  we always  find flow profiles with continuously varying gradients of fluid velocity  for all values of macroscopic strain-rate $\dot\gamma$ implying that the picture of two bands of constant strain rate would be at best an idealisation.  

(ii) An alternative scenario that is observed in other driven systems, such as charge density waves in anisotropic metals~\cite{Maeda1990} and collections of motor proteins~\cite{JulicherProst1995}, is shown in the bottom right frame of Fig.~\ref{fig:scenarios}. In this case the system is expected to exhibit hysteresis, with regions that accommodate coexistence of a range of macroscopic strain rates, corresponding to the constant value $\pm\sigma_0$ of applied stress. In general $\sigma_0$ may coincide with $\sigma_m$ or may be lower, with the system exhibiting ``early swtching". The width of the horizontal hysteretic region of the stress-strain curve decreases with increasing $\alpha$. In this picture the particular steady-state behaviour observed will depend on the  initial conditions and particular flow history of each sample.

(iii)  Another possibility  is  that the system shows a yield-stress like behaviour with a yield stress $\pm \sigma_y$  whose sign is determined by the direction of the flow. The value of the yield stress could also be anywhere in the ``unstable'' range of stress:  $\sigma_y \le \sigma_m$.

(iv) Finally, there is one more possibility:  that he theoretical curve would indeed be reproduced by an experiment which scanned through different values of the macroscopic strain rate. The theoretical curve has been calculated by fixing $\dot\gamma$ and calculating the corresponding value of $\sigma$  under the assumption that there are variations in the director and flow field {\em only in} the gradient direction (i.e. perpendicular to the plates). If this {\em assumption} is valid, every point on this curve does therefore represent a stable state corresponding to this procedure. 

\begin{figure}[t]
\centering
\includegraphics[width=.9\columnwidth]{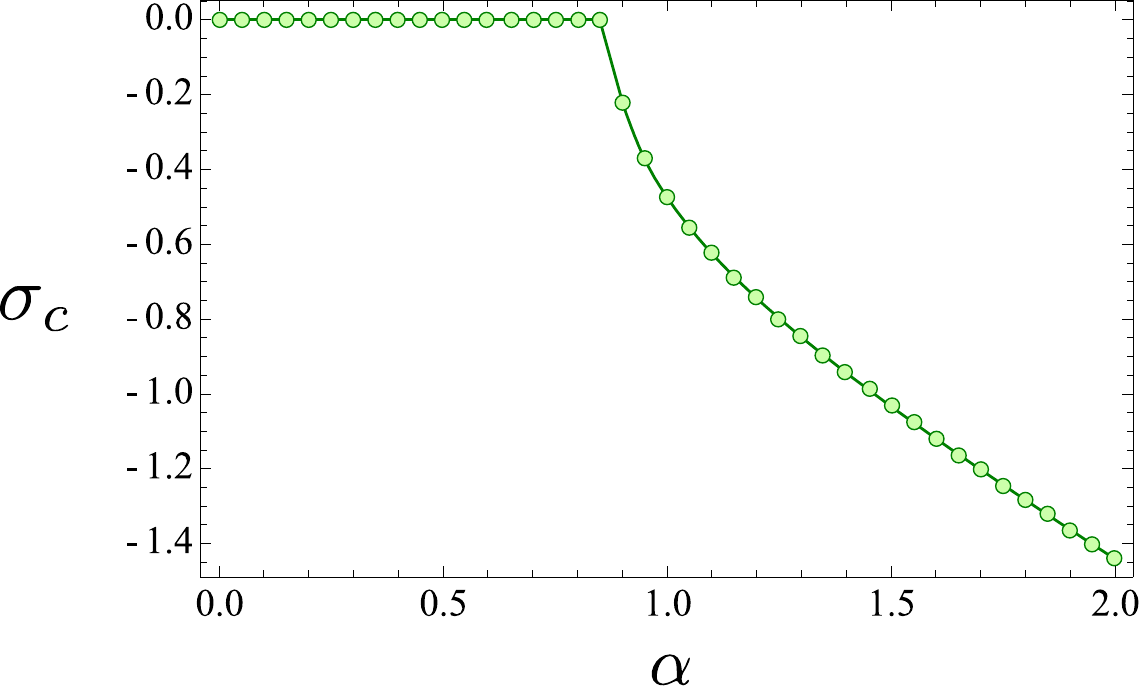}
\caption{\label{fig:yieldstress} Yield-stress $\sigma_{c}$ as a function of $\alpha$ for a nematic suspension ($\beta=w=0$) obtained by numerical solution of the active hydrodynamic equations.}
\end{figure}

For $\alpha>\alpha_{c2}$ the stress-strain curve intercepts the $\dot\gamma=0$ axis at a finite value $\sigma_c=\sigma(\dot\gamma=0)$ of the strain rate. The active suspension has a nonzero spontaneous stress even in the absence of applied forces, as indeed observed in the spontaneous flow regime of an active suspension confined between two stationary no-slip planes.  In other words, a finite force must be applied to the active suspension to keep it from sliding even at zero mean strain rate. This spontaneous stress $\sigma_c$ is shown as a function of $\alpha$ in  Fig.~\ref{fig:yieldstress}.  The sign of the stress determines the direction of spontaneous flow. 

\begin{figure}[t]
\centering
\includegraphics[width=1.\columnwidth]{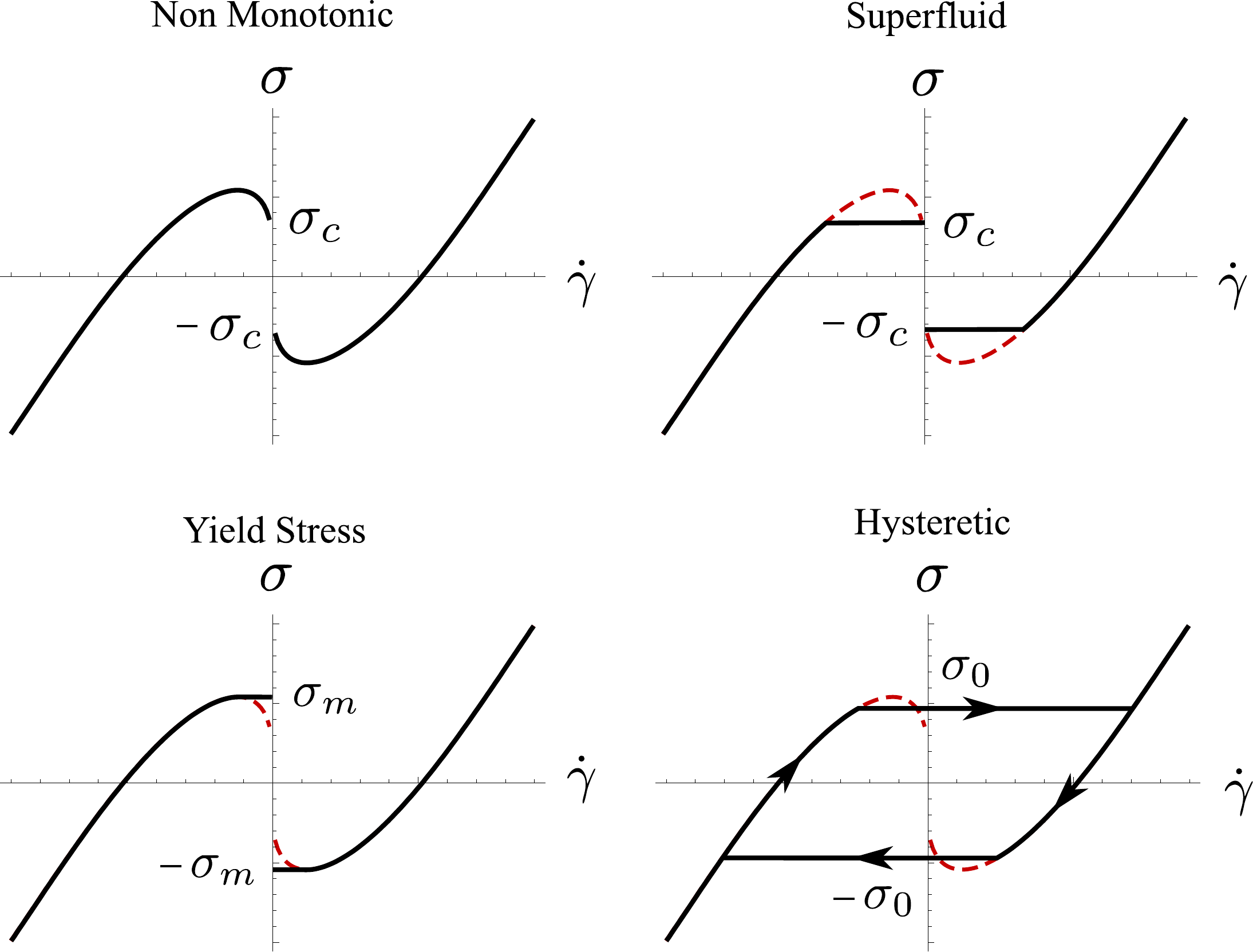}
\caption{\protect
Possible scenarios for the transition to the yield stress regime at $\alpha>\alpha_{c2}$. The non-monotonic curve obtained numerically is shown in the top left frame. In the superfluid scenario (top-right) the plateau at $\sigma=0$ divides into two disconnected branches terminating at $\sigma=\pm \sigma_{c}$. In this case the yield stress is expected to grow monotonically from zero. In the yield stress scenario (bottom-left), there is already a non-zero stress at $\dot\gamma=0$ and thus the yield stress simply continues increasing with no qualitative change in the behaviour at $\alpha_{c2}$. In the hysteretic scenario (bottom-right), the loop intersect the positive $\sigma$ axis at $\pm\sigma_0$, with  $\sigma_c\leq\sigma_0\leq\sigma_m$.}
\label{fig:scenarios2} 
\end{figure}

We now speculate on the possible behavior of the system for each of the scenarios sketched above as $\alpha$ goes through $\alpha_{c2}$. The behavior is shown schematically in Fig.~\ref{fig:scenarios2}. 
(i) In the superfluid scenario, the response of the suspension to an applied macroscopic strain rate will show yield stress behavior. The system would smoothly go from the zero-stress plateau to a yield stress which increases from zero at $\alpha_{c2}$.
(ii) In the hysteretic scenario the minimum height of the hysteretic loop  becomes $2 \sigma_c$ i.e. $\sigma_c \le \sigma_0 \le \sigma_m$. 
(iii) In the yield-stress scenario the system already shows yield stress behaviour which continues for $\alpha > \alpha_{c2}$.
(iv) In the non-monotonic scenario, the non-monotonic stress-strain rate curve shows a jump at $\dot\gamma$ whose magnitude increases from zero at $\alpha_{c2}$.

\section{Discussion and conclusions}

We have studied the rheological behavior of a thin film of polar and apolar active  material. For weakly active systems,  in the regime of the linear rheology, we have confirmed analytically the prediction of Hatwalne and collaborators~\cite{Hatwalne04}  that activity can  lower the linear bulk viscosity of tensile suspensions of swimmers as well as enhance the viscosity of contractile  systems. We have shown that this result applies also for finite systems, in the presence of boundaries. 

An important new result of our work is the role of the {\em shape} of the active particles in controlling the rheological behavior. We find  a remarkable exact duality  that holds in the regime where the stress-strain rate relation is linear and shows that tensile ($\alpha<0$) rod-shaped flow-aligning particles ($\lambda>1$) are rheologically equivalent to contractile ($\alpha>0$) discotic flow-tumbling particles ($-1\leq\lambda<0$). This means that activity lowers the linear viscosity of  both tensile, rod shaped particle and contractile, disc shaped particle suspensions, while it increases the linear viscosity of contractile, rod-shaped particle and tensile, discotic particle suspensions.

For strongly active systems we find that the rheological response is intrinisically nonlinear. The  regime of linear rheology at small strain rates vanishes beyond a critical value of activity.  In this strongly active regime, we explore a number of possible scenarios for the nonlinear rheology which include a ``superfluid'' phase with vanishing viscosity, hysteresis, yield-stress behavior and non-monotonic behavior. Our one-dimensional analysis does not, however, allow us  to determine which of these scenarios is more likely. It is of course possible that allowing for variations of the director and flow field in higher dimensions or allowing for variations in the magnitude of the order parameter would yield a criterion for selecting one of the proposed scenarios.

\begin{acknowledgements}
LG is supported by NSF through the Harvard MRSEC and the Brandeis MRSEC and by the Harvard Kavli Institute for Nanobio Science \& Technology. MCM is supported by NSF grants DMR-075105 and DMR-0806511. TBL acknowledges the support of  EPSRC under grant EP/G026440/1. We thank Suzanne Fielding and James Adams for illuminating discussions.
\end{acknowledgements}

\appendix

\section{\label{sec:appendix}Derivation of Eqs. (3)}

In this section we show some details of the derivation of the modified ``passive'' terms in the  equation for the director field ${\bf p}$ in the polarized state, when fluctuations in the magnitude of the order parameters are neglected. The equation for the full vector order parameter ${\bf P}$ has the form
\begin{equation}\label{eq:ap1}
[\partial_{t}+{\bf v}\cdot\nabla]P_{i} = \lambda u_{ij}P_{j} + \Gamma h_{i}+\Gamma'f_{i}\,.	
\end{equation}
Eq. \eqref{eq:ap1} can be separated in two equations for the magnitude $P=|{\bf P}|$ of the polarization and its direction ${\bf p}={\bf P}/P$, using
\begin{gather}
\partial_{t} P = p_{i}\partial_{t}P_{i}\,,\\[5pt]
\partial_{t} p_{i} = \frac{1}{P}\,\delta_{ij}^{T}\partial_{t}P_{j}\,,	
\end{gather}
where $\delta_{ij}^{T}=\delta_{ij}-p_{i}p_{j}$ is a transverse projection operator, with the result
\begin{gather*}
\partial_{t}P = P (\lambda u_{ij}p_{i}p_{j})+\Gamma'f_{\parallel}+\Gamma h_{\parallel}\,,\\
[\partial_{t}+{\bf v}\cdot\nabla]p_{i}+\omega_{ij}p_{j} = \lambda\delta_{ij}^{T}u_{jk}p_{k}+\frac{1}{P}(\Gamma'f_{i}^{\perp}+\Gamma h_{i}^{\perp})\,,
\end{gather*}
where we have defined
\[
h_{\parallel} = {\bf p}\cdot{\bf h}\,, \quad
h_{i}^{\perp} = \delta_{ij}^{T}h_{j}\,,\quad 
f_{\parallel} = {\bf p}\cdot{\bf f}\,, \quad
f_{i}^{\perp} = \delta_{ij}^{T}f_{j}\,.
\]
In the ordered state, fluctuations in the magnitude $P$ of the polarization are overdamped and will be neglected. We can assume, on the other hand, to be deeply in the polarized state and that $P=\sqrt{-a_{2}/a_{4}}$ is constant. For simplicity we redefine the units so that $P=1$. The condition $P={\rm const}$ determines the longitudinal part $h_{\parallel}$ of the molecular field. This requires
\[
h_{\parallel}=-\frac{1}{\Gamma}[\Gamma'f_{\parallel}+\lambda u_{ij}p_{i}p_{j}]\,.
\]
The above expression can be now used to eliminate $h_{\parallel}$ from the density ${\bf j}=\Gamma' {\bf h}+\Gamma''{\bf f}$ appearing at the right-hand side of Eq. \eqref{eq:c1}. Expressing $h_{i}=p_{i}h_{\parallel}+h_{i}^{\perp}$ and $f_{i}=p_{i}f_{\parallel}+f_{i}^{\perp}$ we obtain
\begin{equation}\label{eq:ap2}
j_{i} = p_{i}\Gamma''(1-\xi)f_{\parallel}-\gamma'\lambda u_{kl}p_{k}p_{l}p_{i}+\Gamma'h_{i}^{\perp}+\Gamma''f_{i}^{\perp}\,,
\end{equation}
where $\xi=(\Gamma')^2/(\Gamma\Gamma'')$ is a dimensionless parameter and $\gamma'=\Gamma'/\Gamma$.
Similarly, the stress tensor $\sigma_{ij}^{r}$ becomes
\begin{multline}\label{eq:ap3}
\sigma_{ij}^{r} = -\delta_{ij}\Pi-\frac{\lambda}{2}[p_{i}h_{j}^{\perp}+p_{j}h_{i}^{\perp}]\\
+\frac{1}{2}[p_{i}h_{j}^{\perp}-p_{j}h_{i}^{\perp}]-\lambda p_{i}p_{j} h_{\parallel}\,.
\end{multline}
The longitudingal and transverse parts of the driving force $f_{i}$ are given by
\begin{gather*}
f_{\parallel} = -\frac{C}{c_{0}^{2}}\,{\bf p}\cdot\nabla c -\frac{B_{1}-B_{3}}{c_{0}}\,{\bf p}\cdot\nabla(\nabla\cdot{\bf p})\,,\\
f_{i}^{\perp} = \delta_{ij}^{T}\left[-\frac{C}{c_{0}^{2}}\,\partial_{j}c-\frac{B_{1}-B_{3}}{c_{0}}\,\partial_{j}\nabla\cdot{\bf p}\right]\,.
\end{gather*}
Similarly, the transverse part of the molecular field is given by
\[
h_{i}^{\perp} = \delta_{ij}^{T}\left[\frac{B_{1}-B_{3}}{c_{0}}\,\partial_{j}c+(K_{1}-K_{3})\partial_{j}\nabla\cdot{\bf p}+K_{3}\nabla^{2}p_{j}\right]\,.
\]
Replacing the explicit expressions of $h_{\parallel}$, $h_{i}^{\perp}$, $f_{\parallel}$ and $f_{i}^{\perp}$ in Eqs. \eqref{eq:ap2} and \eqref{eq:ap3}, we finally obtain
\[
j_{i} = 
-\left[D\left(1-\xi\right)p_ip_j-\frac{\gamma'w}{c_0}\delta_{ij}^T\right]\partial_{j}c
-\gamma'\lambda u_{kl}p_{k}p_{l}p_{i}\,,
\]
where $w=\Gamma(B_1-B_3)$ is a velocity and we have neglected terms of second and higher order in the hydrodynamic fields. Finally, the reversible part of the stress tensor is given by
\begin{widetext}
\begin{align*}
\label{eq:sigmafull}
\sigma_{ij}^r 
&= -\delta_{ij}\Pi + \lambda p_{i}p_{j}p_{k}\left[\frac{B_{1}-B_{3}}{c_{0}}\,\partial_{k}c+(K_{1}-K_{3})\partial_{k}\nabla\cdot{\bf p}+K_{3}\nabla^{2}p_{k}\right]\\[5pt]
&-\frac{\lambda}{2}\left[\frac{B_{1}-B_{3}}{c_{0}}(p_{i}\partial_{j}c+p_{j}\partial_{i}c)
+(K_{1}-K_{3})(p_{i}\partial_{j}+p_{j}\partial_{i})\nabla\cdot{\bf p}
+K_{3}(p_{i}\nabla^{2}p_{j}+p_{j}\nabla^{2}p_{i})\right]\\[2pt]
&+\frac{1}{2}\left[\frac{B_{1}-B_{3}}{c_{0}}(p_{i}\partial_{j}c-p_{j}\partial_{i}c)
+(K_{1}-K_{3})(p_{i}\partial_{j}-p_{j}\partial_{i})\nabla\cdot{\bf p}
+K_{3}(p_{i}\nabla^{2}p_{j}-p_{j}\nabla^{2}p_{i})\right]\\
&-\lambda\Gamma'\xi\,p_{i}p_{j}(Dp_{k}\partial_{k}c+wp_{k}\partial_{k}\partial_{l}p_{l})
+\frac{\lambda^{2}}{\Gamma} p_{i}p_{j} u_{kl} p_{k}p_{l}\;.
\end{align*}
\end{widetext}
Taking $K_{1}=K_{3}=K$  leads to the equations given in Sec. \ref{sec:introduction}.


\begin{thebibliography}{99}
\bibitem{defnote}
The  parameters $\beta_i$ and $\alpha$ have different dimensions  from those used in~\cite{TBLMCMbook} as here we have incorporated an additional factor of $\ell^2$ in their definition.
\end{thebibliography}
\end{document}